\documentclass[12pt]{iopart}

\usepackage{iopams, graphicx}

\usepackage[svgnames]{xcolor}

\definecolor{darkred}{rgb}{0.5,0,0}
\definecolor{darkgreen}{rgb}{0,0.5,0}
\definecolor{darkblue}{rgb}{0,0,0.5}

\begin{document}

\title[Signatures of quantum mechanics in chaotic systems]{Signatures of quantum mechanics in chaotic systems}

\author{Kevin M Short$^1$ and Matthew A Morena$^2$}

\address{$^1$ Integrated Applied Mathematics Program, University of New Hampshire, Durham, NH 03824}
\address{$^2$ Division of Mathematics and Science, Young Harris College, Young Harris, GA 30582}
\ead{Kevin.Short@unh.edu}

\vspace{10pt}
\begin{indented}
	\item[] Received 24 October 2016
\end{indented}

\begin{abstract}
We consider the quantum-classical correspondence from a classical perspective by discussing the potential for chaotic systems to support behaviors
normally associated with quantum mechanical systems. Our main analytical tool is a chaotic system's set of cupolets, which are essentially
highly-accurate stabilizations of its unstable periodic orbits. The discussion is motivated by the bound or entangled states that we have recently
detected between interacting chaotic systems, wherein pairs of cupolets are induced into a state of mutually-sustaining stabilization that can be
maintained without external intervention. This state is known as chaotic entanglement as it has been shown to exhibit several properties consistent
with quantum entanglement. For instance, should the interaction be disturbed, then the chaotic entanglement would be broken. In this paper, we further
describe chaotic entanglement and go on to discuss the capacity for chaotic systems to exhibit other characteristics that are conventionally
associated with quantum mechanics, namely analogs to wave function collapse, the measurement problem, the superposition of states, and to quantum
entropy definitions. In doing so, we argue that these characteristics need not be regarded exclusively as quantum mechanical.
\end{abstract}

\vspace{2pc}


\submitto{\NL}

\noindent (Some figures may appear in colour only in the online journal)


\section{Introduction}
\label{sec:introduction}

Chaotic behavior is generally attributed to a sensitive dependence on initial conditions and is characterized by a positive maximal Lyapunov exponent
that causes nearby trajectories to diverge from each other exponentially fast. Despite its ubiquity in classical physics, chaos is yet to be rigorously
established within quantum settings. One explanation for this disparity is that unlike chaotic or classical systems, whose states may be completely
described by a set of dynamical variables, in quantum mechanics conjugate observables such as position and momentum cannot take on well-defined values
at the same time. Particle dynamics are instead determined in part by the uncertainity principle and by the linearity of the Schr\"{o}dinger equation,
which preserves the overlap between quantum states. In other words, the nonlinearity required for chaotic dynamics and the exponential divergence of
neighboring trajectories seem fundamentally incompatible with quantum mechanics in its present formulation.

And yet, much effort has recently been devoted to detecting signatures of chaos in quantum
systems~\cite{Habib1998,Nielsen2000,Haake2001,Ghose2008b,Chaudhury2009}. One such signature is the sensitivity of some quantum systems to
perturbation. This has been experimentally observed in the decay in the overlap between quantum states that are evolving under slightly different
Hamiltonians and is thought to be associated with a positive Lypaunov exponent~\cite{Habib1998,Ghose2008b}. In particular, the rate of overlap decay
is known to transpire at different rates depending on whether the evolution begins from initial conditions that correspond classically to chaotic
versus regular regimes~\cite{Chaudhury2009}. A second signature is quantum scarring, which refers to the scenario in which a quantum system's
associated wave function is concentrated on paths that represent periodic orbits in the classical limit~\cite{Heller1984,Berry1989b}.
This phenomenon has been experimentally observed in several recent studies~\cite{Doya2001,Michel2007}. 

Entanglement in the purely quantum sense has also been observed to be a reliable indicator of classical
chaos~\cite{Furuya1998,Fujisaki2003,Wang2004,McHarris2011}. For example, in Chaudhury \emph{et al}.'s recent kicked top experiments of laser-cooled
Cesium ($^{133}$Cs) atoms, each atom's initial state is followed for several periods of the ``kicked'' Hamiltonian, and the corresponding classical
phase space reveals islands of regular motion surrounded by a sea of chaos~\cite{Chaudhury2009}. When entropy is used to measure the entanglement,
greater entanglement is detected in initial states that are prepared from chaotic regimes as opposed to weaker entanglement which is generated by those
originating from regular regions. It is as if the quantum regime respects an underlying classical presence~\cite{McHarris2011}.

One feature of chaotic systems typically encountered in investigations is an infinite set of unstable periodic orbits (UPOs) that are found densely
embedded in many attractors. These orbits collectively provide a rich source of qualitative information about the parent chaotic system and are a
focus of numerous theoretical and practical applications~\cite{Franzosi2005,Short2005a,Short2005b}. As a result, several control schemes have been
designed to detect and stabilize these orbits~\cite{Auerbach1987,Ott1990,Hayes1993,Hayes1994}. In Section~\ref{sec:cupolet_background}, we discuss an
adaptation of one particular control method that very efficiently stabilizes the UPOs of chaotic systems onto \emph{cupolets}
\emph{(\underline{C}haotic, \underline{U}nstable, \underline{P}eriodic,
\underline{O}rbit-\underline{LETS})}~\cite{Parker1999,Short1999,Zarringhalam2007,Morena2014c}. Cupolets are controlled and stabilized periodic orbits
of a chaotic system that would normally be unstable without the presence of the control mechanism. These orbits represent a subset of the UPOs, but
are distinguished because this stabilization supports a one-to-one correspondence between a given sequence of controls and a specific cupolet, with
each cupolet able to be generated independently of initial condition. All of this allows for large collections of cupolets to be generated very
efficiently, thereby making these orbits well suited for analyzing chaotic systems.

In recent studies, we reported on the proclivity for chaotic systems to enter into bound or entangled
states~\cite{Morena2013,Morena2014b,Morena2014c}. We demonstrated how pairs of interacting cupolets may be induced into a state of mutually-sustaining
stabilization that requires no external controls in order to be maintained. This state is known as \emph{chaotic entanglement} and it is
self-perpetuating within the cupolet-stabilizing control scheme, meaning that each cupolet of an entangled pair is effectively controlling the
stability of its partner cupolet via their continued interaction. The controls used are all information-theoretic, so we stress that additional work
is required to more rigorously relate this research to physical systems. However, since many of our simulated cupolet-to-cupolet interactions are
based on the dynamics of physical systems, our findings indicate the potential of chaotic entanglement to be both physically realizable and naturally
occurring. It is worth noting the sensitivity of chaotic entanglement to disturbance since any disruption to the stability of either cupolet of an
entangled pair may be enough to destroy the entanglement, therefore supporting a reasonable analog to quantum entanglement.

We are aware that entanglement is regarded as a quantum phenomenon and that there are characteristics of quantum entanglement that are not compatible
with chaotic entanglement, such as nonlocality. We are also aware that chaotic entanglement has been previously examined in~\cite{Zhang2013} and that
a classical version of entanglement has been proposed in~\cite{Spreeuw1998}. In the first study, linear and nonlinear subsystems are coupled together
to produce composite chaotic systems, a synthesis the authors refer to as chaotic entanglement. In the second study, a classical version of quantum
entanglement is demonstrated via a beam of photons and is shown to be consistent with many features of quantum entanglement, apart from nonlocality.
In contrast, the novelty of the chaotic entanglement that we have documented arises in how two interacting chaotic systems are induced into a state of
mutual stabilization. First, the chaotic behavior of the two systems is collapsed onto unique periodic orbits (cupolets). Following the collapse, the
ensuing periodicity of each chaotic system and the stability of each cupolet are maintained intrinsically by each system's dynamical behavior and will
persist until the interaction is disturbed. To our knowledge, this is the first documentation of chaotic systems interacting to such an extent.

Our initial results are very promising since several hundred pairs of entangled cupolets have been identified from low-dimensional chaotic systems.
When regarded as a parallel to quantum entanglement, chaotic entanglement is further intriguing as it not only signals a new correlation between
classical and quantum mechanics, but it also demonstrates that chaotic systems are capable of exhibiting behavior that has conventionally been
associated with quantum systems. We now discuss the potential for classically chaotic systems to support additional parallels with quantum mechanics,
namely the measurement problem, notions of wave function collapse, superposition of states, and entropy definitions.

Our discussion uses cupolets and chaotic entanglement as reference points and is organized as follows. In Section~\ref{sec:cupolet_background}, we
begin by providing a brief introduction to cupolets and how they are generated, and then we discuss a few of their interesting properties and
applications. In Section~\ref{sec:chaotic_entanglement}, we describe chaotic entanglement and how it can be induced and detected between pairs of
interacting cupolets. The main discussion of chaotic systems supporting quantum behavior is found in Section~\ref{sec:main_discussion}. Finally, we
offer a few concluding remarks in Section~\ref{sec:concluding_remarks}.



\section{Background on cupolets}
\label{sec:cupolet_background}

Broadly speaking, cupolets are a relatively new class of waveforms that were originally detected while controlling a chaotic system in a secure
communication application. The theory behind these orbits and their applications have been
well-documented~\cite{Parker1999,Short1999,Short2005a,Short2005b,Zarringhalam2007,Zarringhalam2008,Morena2013,Morena2014a,Morena2014b,Morena2014c}. In
this section, we summarize the control technique that is used to generate cupolets and then describe the applications of cupolets that have particular
relevance to our chaotic entanglement research. More technical details of the control process can be found
in~\cite{Zarringhalam2007,Zarringhalam2008,Morena2014a,Morena2014b,Morena2014c}.

\begin{figure}[!t]
	\centering
	\includegraphics[scale=0.425]{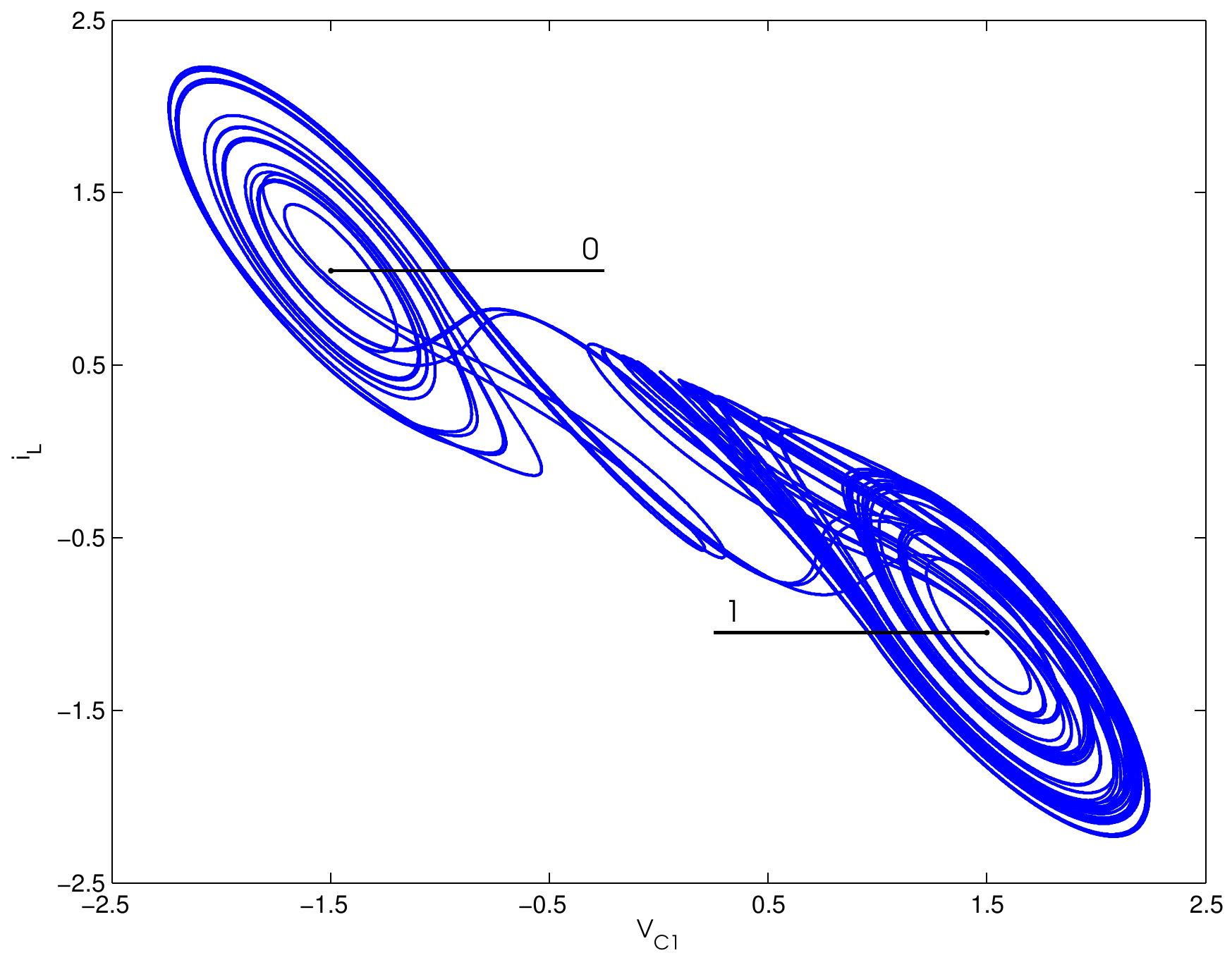}
	\caption{2D projection of the double scroll attractor showing the control surfaces~\cite{Matsumoto1985}.{\label{fig:DS_attractor}}}
\end{figure}

The control scheme that is used to stabilize cupolets is adapted from the chaos control method designed by Hayes, Grebogi, and Ott
(HGO)~\cite{Hayes1993,Hayes1994}. In the HGO scheme, small perturbations are used to steer trajectories of the double scroll system, also known as
Chua's oscillator~\cite{Matsumoto1985}, around an attractor. The differential equations describing this system are given by:
\begin{eqnarray}
	\dot{v}_{C_{1}} &= \frac{G(v_{C_{2}}-v_{C_{1}})-g(v_{C_{1}})}{C_{1}}, \nonumber  \\
	\dot{v}_{C_{2}} &= \frac{G(v_{C_{1}}-v_{C_{2}})+i_{L}}{C_{2}},\\
	\dot{i}_{L} &= -\frac{v_{C_{2}}}{L}, \nonumber
\end{eqnarray}
where the piecewise linear function, $g(v)$, is given by:
\begin{eqnarray}
	g(v) &= \left\{
	\begin{array}{l}
		m_{1}v, \\
		m_{0}\left( v+B_{p} \right) - m_{1}B_{p}, \\
		m_{0}\left( v-B_{p} \right) + m_{1}B_{p},
	\end{array}
	\begin{array}{r}
		\textrm{if} \\
		\textrm{if} \\
		\textrm{if}
	\end{array}
	\begin{array}{l}
		\vert v \vert \leq B_{p}, \\
		v \leq -B_{p}, \\
		v \geq B_{p}.
	\end{array}
	\right.
\end{eqnarray}
When $C_{1}=\frac{1}{9}$, $C_{2}=1$, $L=\frac{1}{7}$, $G=0.7$, $m_{0}=-0.5$, $m_{1}=-0.8$, and $B_{p}=1$, the double scroll system is known to be
chaotic and its attractor consists of two lobes that each surrounds an unstable fixed point~\cite{Matsumoto1985}. Figure~\ref{fig:DS_attractor} shows a
typical trajectory tracing out this attractor.

Control of the double scroll system is first achieved by setting up two control planes on the attractor (via a Poincar\'{e} surface of section) and
then by partitioning each control plane into small control bins. Perturbations are applied only when a trajectory evolves through the control bins,
otherwise the trajectory is allowed to freely evolve around the attractor. Figure~\ref{fig:DS_attractor} also shows the positions of these control
planes which emanate outward from the center of each lobe. The control planes are assigned binary values so that a binary symbolic sequence may be
recorded whenever a trajectory intersects a control plane. This sequence is known as a \emph{visitation sequence}.

\begin{figure*}[!t]
	\centering
	\begin{tabular}{cc} 
		\includegraphics[width=0.35\columnwidth]{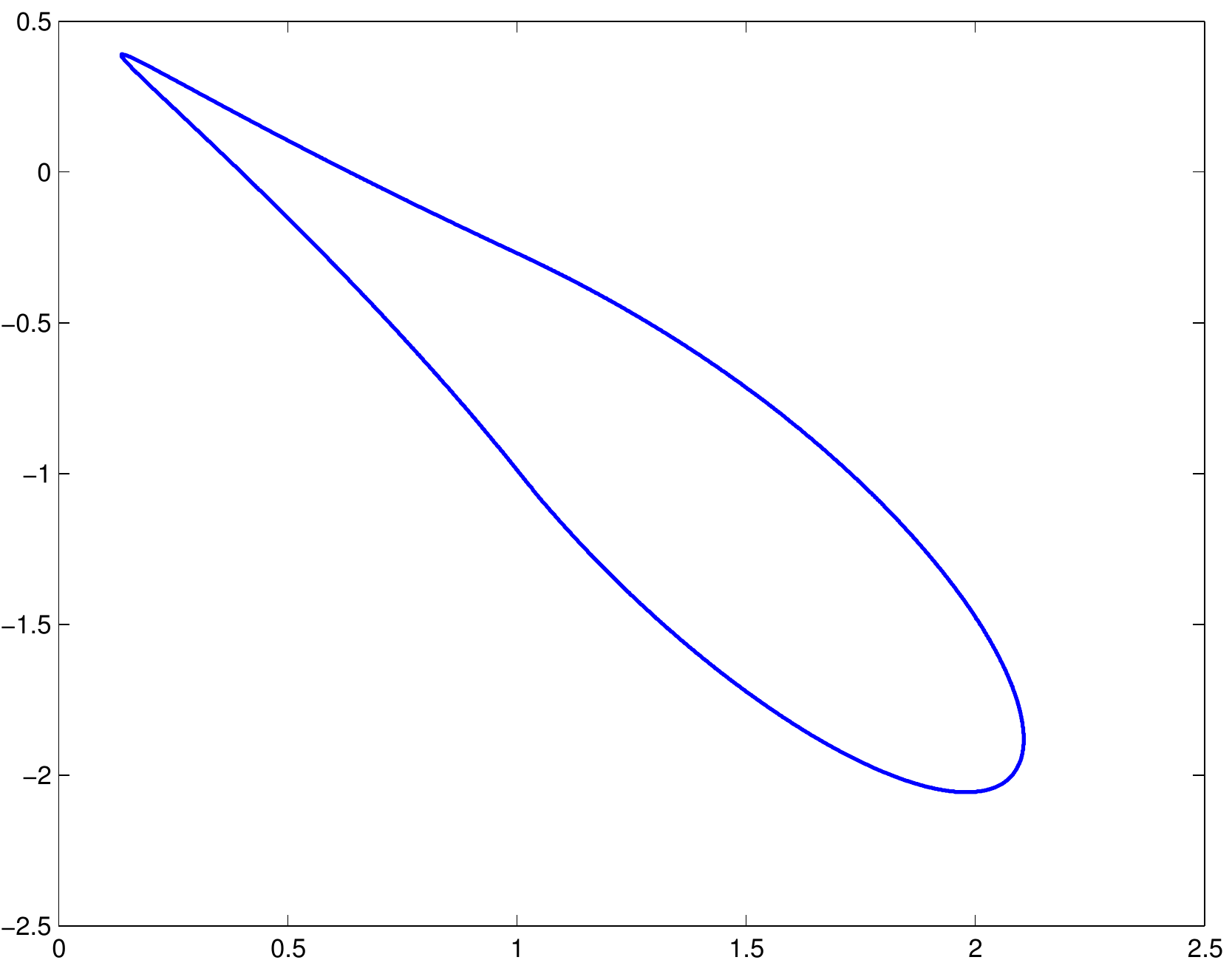} &
		\includegraphics[width=0.35\columnwidth]{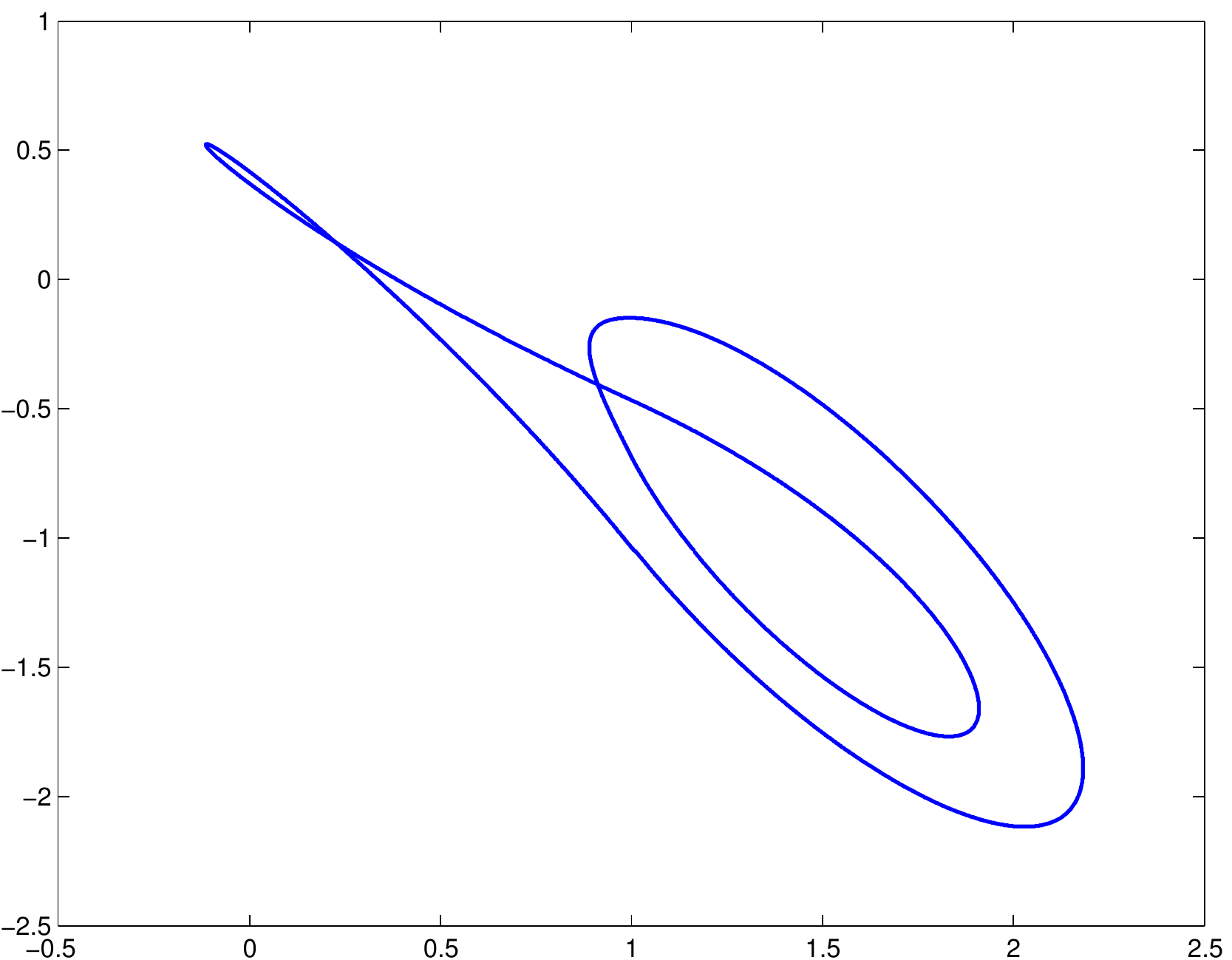} \\
		\small{(a)} & \small{(b)} \\ 
		\includegraphics[width=0.35\columnwidth]{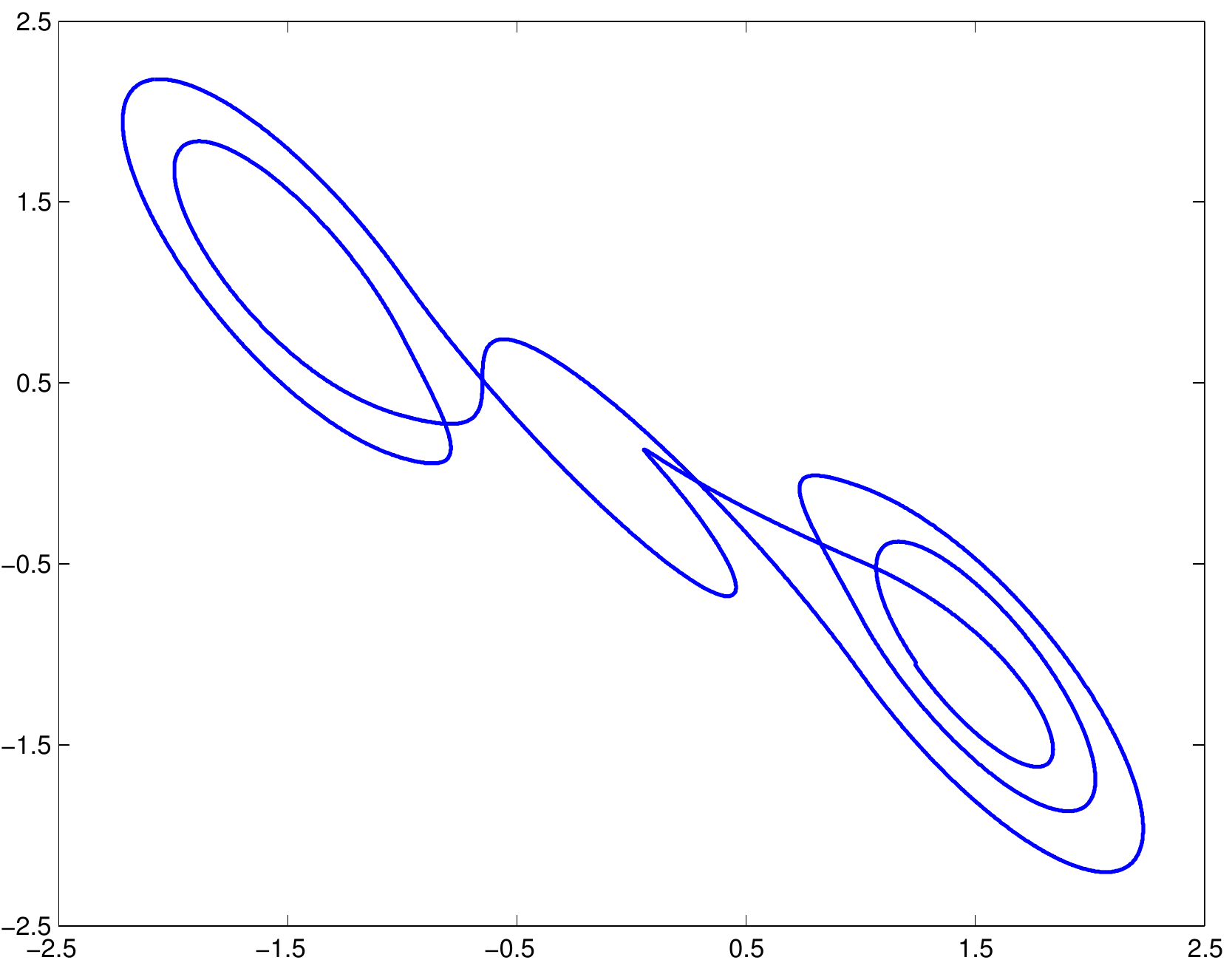} &
		\includegraphics[width=0.35\columnwidth]{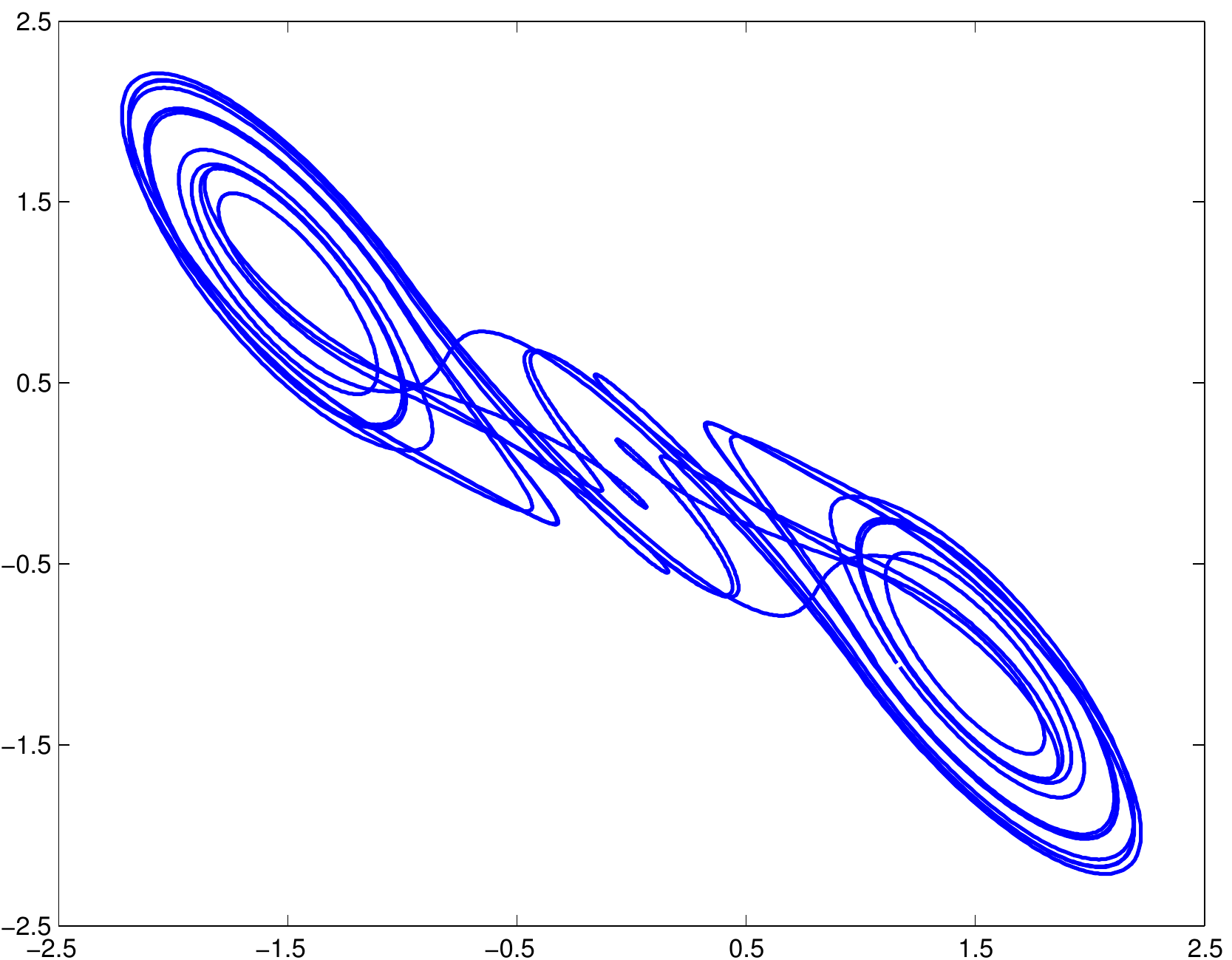} \\
		\small{(c)} & \small{(d)}
	\end{tabular} 
	\caption{Cupolets of various periods belonging to the double scroll system. The control sequences which must be periodically applied in order to stabilize
	their periodic orbits are (a) `$00$', (b) `$11$', (c) `$00001$', and (d) `$001$'~\cite{Morena2014b}.{\label{fig:four-cupolets}}} 
\end{figure*}

Parker and Short~\cite{Parker1999} later combined this control scheme with ideas from the study of impulsive differential equations~\cite{Bainov1989}
and discovered that when a repeating binary \emph{control sequence} is used to define the controls, with a `1' bit corresponding to a perturbation and
a `0' bit corresponding to no perturbation, then the double scroll system stabilizes onto a periodic orbit. These perturbations are defined via the HGO
technique to be the smallest disturbance along a control plane that produces a change of lobe $M$ loops downstream. For almost all repeating control
sequences, the resulting periodic orbits are generated completely independently of the initial state of the system and a one-to-one relation exists
between a given control sequence and a particular periodic orbit. These periodic orbits have been given the name \emph{cupolets}, and this work has
since been extended to chaotic maps and a variety of other continuous chaotic systems such as the Lorenz and R\"{o}ssler systems. The examples of
double scroll cupolets appearing in Figure~\ref{fig:four-cupolets} are generated by repeating the indicated sequences of control bits.

To summarize, cupolets are highly-accurate approximations to the UPOs of chaotic systems that are generated by an adaptation of the HGO control
technique~\cite{Parker1999,Zarringhalam2007,Morena2014c}. Cupolets have the interesting properties of being stabilized independently of initial
condition and also of being in one-to-one correspondence with the control sequences. These controls can be made arbitrarily small and thus do not
significantly alter the topology of the orbits on the chaotic attractor. This suggests that cupolets are shadowing true periodic orbits and theorems
have been developed to establish conditions under which this holds~\cite{Grebogi1990,Sauer1991,Coomes1997,Hammel1988,Zarringhalam2007}. Furthermore,
the effect of combining chaos control with impulsive perturbations has resulted not only in the ability to stabilize chaotic systems onto (approximate)
periodic orbits, but has also simplified the search for periodic orbits since a simple program can be written to generate all possible N-bit control
sequences and then feed them into the control scheme. What further distinguishes cupolets from UPOs, which are traditionally stabilized via techniques
such as Newton's or first-return algorithms, is that large numbers of cupolets can be inexpensively generated by only a few bits of binary control
information. For example, over 8,800 double scroll cupolets can be stabilized from $16$-bit or fewer control sequences.

\subsection{Application of Cupolets}
\label{sub:application_of_cupolets}

\begin{figure*}[!t]
	\centering
	\begin{tabular}{cc}
		\includegraphics[width=0.475\columnwidth]{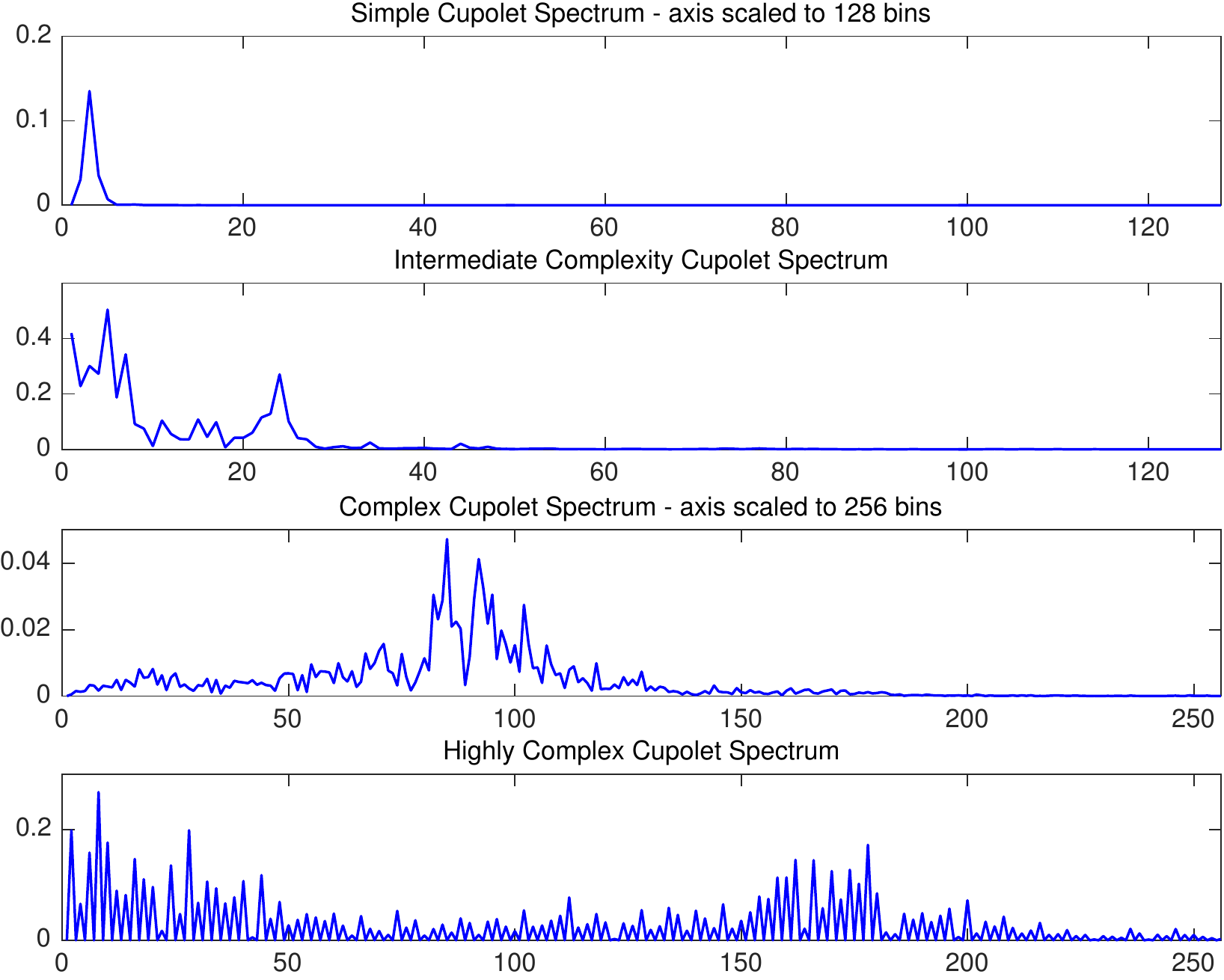} &
		\includegraphics[width=0.475\columnwidth]{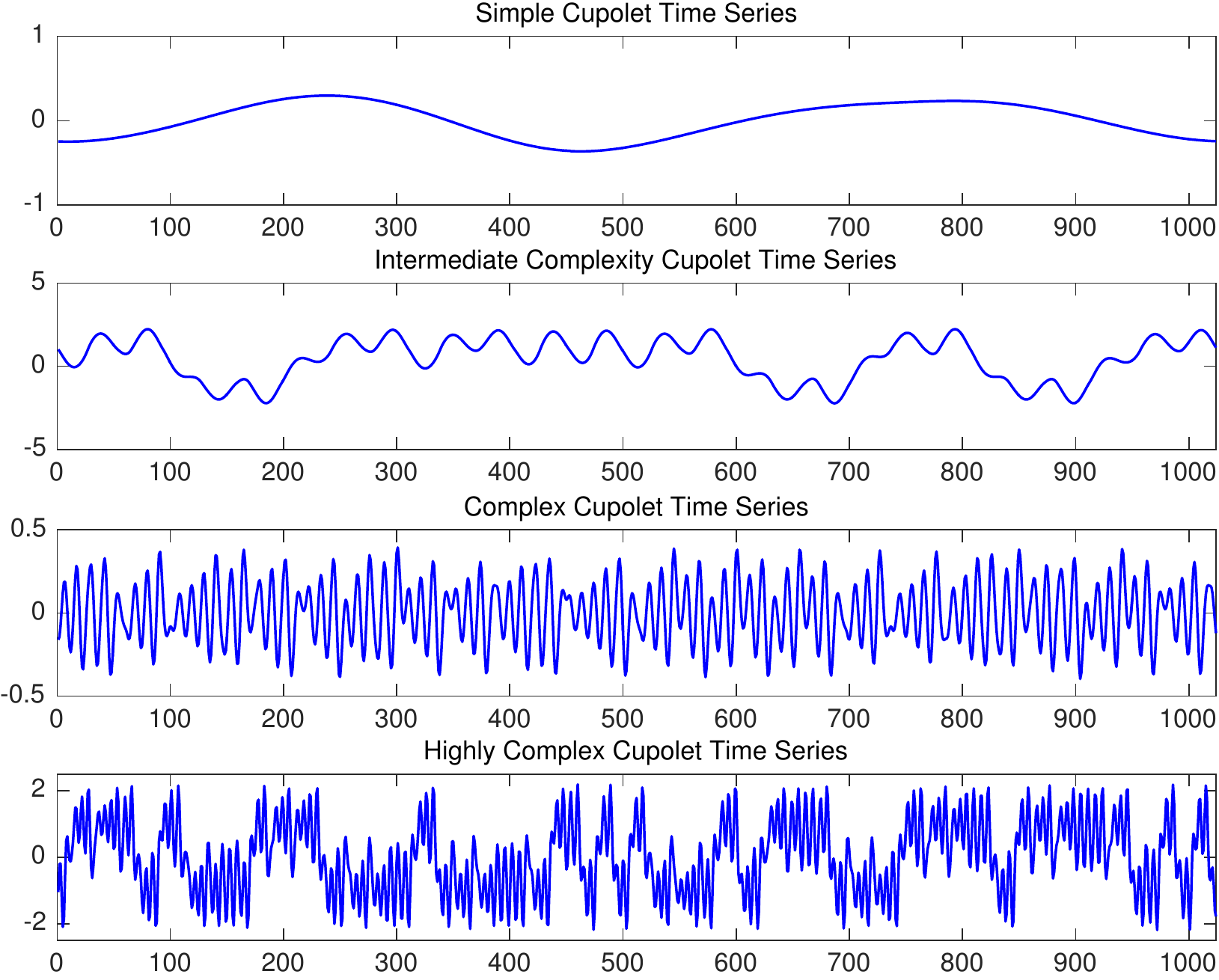} \\[1ex]
		\small{(a)} & \small{(b)}
	\end{tabular}
	\caption{Cupolet diversity: in~(a), spectral variation and in~(b), time-domain variation among cupolets~\cite{Short2005a,Short2005b}. The same cupolets are
	used to produce the corresponding graphs between the two sets of figures.\label{fig:cupolet_diversity}}
\end{figure*}

At a fundamental level, cupolets are very rich in structure and may be used to generate a variety of different waveforms ranging from a simple
\emph{sine}-like wave with a single dominant spectral peak to more involved waveforms consisting of many harmonics.
Figure~\ref{fig:cupolet_diversity}(a) illustrates the high diversity in spectral signatures found among four cupolets. The data is taken from the FFT
of a single period of oscillation of each cupolet, and each cupolet's corresponding time domain representation is seen in
Figure~\ref{fig:cupolet_diversity}(b). It is clear that the simplest cupolet in these figures is essentially sinusoidal, while increasingly richer
structure is evident in the other cupolets. This figure will be referenced later in Section~\ref{sec:main_discussion}.

In addition to secure communication~\cite{Parker1999,Short1999}, image processing~\cite{Zarringhalam2008}, and data
compression~\cite{Short2005a,Short2005b}, cupolets have also been found to provide particularly useful pathways when directing a chaotic system to a
target state~\cite{Morena2014a}. This is a recent application of targeting in dynamical systems and it relies heavily on the fact that cupolet
stabilization occurs independently of the current state of the system. For a given cupolet to remain stabilized, all that is required is the repeated
application of its control sequence to the system, and so applying different controls would induce the system to destabilize from the stabilized
cupolet and to revert to chaotic behavior. If a second sequence of controls were to then be periodically applied, the chaotic system would eventually
restabilize onto a second cupolet, possibly after some intermediary transient phase. Any transient is the result of the trajectory evolving while the
chaotic system sifts through all possible states until it reaches one where the behavior of an UPO falls into synchrony with the control sequence,
thus stabilizing the cupolet. Cupolet restabilization is guaranteed because of the injective relationship that exists between cupolets and the control
sequences. This makes it possible to transition between cupolets, and thus between UPOs, simply by switching control sequences. It is further shown
in~\cite{Morena2014a} that this simple targeting method can be combined with algebraic graph theory and Dijkstra's shortest path algorithm in order to
achieve highly efficient targeting of desired cupolets. We shall refer to cupolet transitions throughout Section~\ref{sec:chaotic_entanglement}.




\section{Chaotic entanglement}
\label{sec:chaotic_entanglement}

In previous work~\cite{Morena2013,Morena2014b,Morena2014c}, we document the surprising observation that pairs of chaotic systems may interact in such a
way that they \emph{chaotically entangle}. To do so, the two chaotic systems must first induce each other to collapse and stabilize onto a cupolet via
the exchange of control information. Then, the stabilities of the two stabilized cupolets must become deterministically linked: disturbing one cupolet
from its periodic orbit subsequently affects the stability of the partner cupolet, and vice versa. Hundreds of entangled cupolet pairs have been
identified for the double scroll system, and it has been shown that chaotic entanglement evokes several connections to quantum entanglement as we
discuss in Section~\ref{sub:chaotic_entanglement_as_an_analog_of_quantum_entanglement}, below.

Cupolets from the two entangled chaotic systems are regarded as mutually stabilizing since their interaction essentially serves as a two-way coupling
that is self-perpetuatating within the control scheme just described in Section~\ref{sec:cupolet_background}. In particular, once entanglement has
been established between two chaotic systems, no outside intervention or user-defined controls are needed to sustain the stabilities of their
respective cupolets. Instead, the stability of each cupolet is maintained by the dynamics of the partner cupolet. In summary, not only has the
original chaotic behavior of the two parent systems collapsed onto the periodic orbits of the two cupolets, but their periodic behavior will persist
as long as their interaction is undisturbed.

Chaotic entanglement is typically mediated by an \emph{exchange function} that defines the interaction between the two chaotic systems and their
cupolets. In~\cite{Morena2014b}, exchange functions are described more fully as catalysts for the entanglement and are taken to represent the
environment or medium in which the chaotic systems are found. For instance, we have designed several types of exchange functions that simulate the
interactions of various physical systems such as the integrate-and-fire dynamics of laser systems and networks of neurons.

\subsection{Chaotic entanglement through cupolets}
\label{sub:chaotic_entanglement_through_cupolets}

In Section~\ref{sec:cupolet_background}, we defined a cupolet's visitation sequence to be the binary sequence of lobes that its orbit visits. Visitation
sequences thus serve as a type of symbolic dynamics of chaotic systems; i.e., dynamic information that is generated as solutions to these systems
evolve over time. With this in mind, chaotic entanglement can be more technically characterized as an exchange of symbolic information in the form of
visitation sequences.

Consider a pair of cupolets, say $\mathbf{C}_{\rm{A}}$ and $\mathbf{C}_{\rm{B}}$, that have been stabilized from two arbitrary but interacting chaotic
systems. As cupolet $\mathbf{C}_{\rm{A}}$ evolves about its attractor, the bits of its visitation sequence are passed to an exchange function which
then performs a binary operation on the visitation sequence. The outputted sequence of bits is known as an \emph{emitted sequence} and is taken as a
control sequence and applied to cupolet $\mathbf{C}_{\rm{B}}$. Concurrently, but in the reverse direction, the visitation sequence belonging to
$\mathbf{C}_{\rm{B}}$ passes through the same exchange function and the resulting emitted sequence is used to control $\mathbf{C}_{\rm{A}}$. At this
point, each cupolet is both receiving and transmitting control information via the exchange function, but if the emitted sequence generated from the
visitation sequence of $\mathbf{C}_{\rm{A}}$ matches the control sequence needed to maintain the stability of cupolet $\mathbf{C}_{\rm{B}}$---and vice
versa---then the two cupolets, and the two parent chaotic systems, become intertwined in a mutually-stabilizing feedback loop and are considered
chaotically entangled. Any external controlling can be subsequently discontinued now that each cupolet's visitation sequence is preserving the partner
cupolet's stabilization.

\begin{figure*}[!t]
	\centering
	\small
	\begin{tabular}{cc} 
		\includegraphics[width=0.425\textwidth]{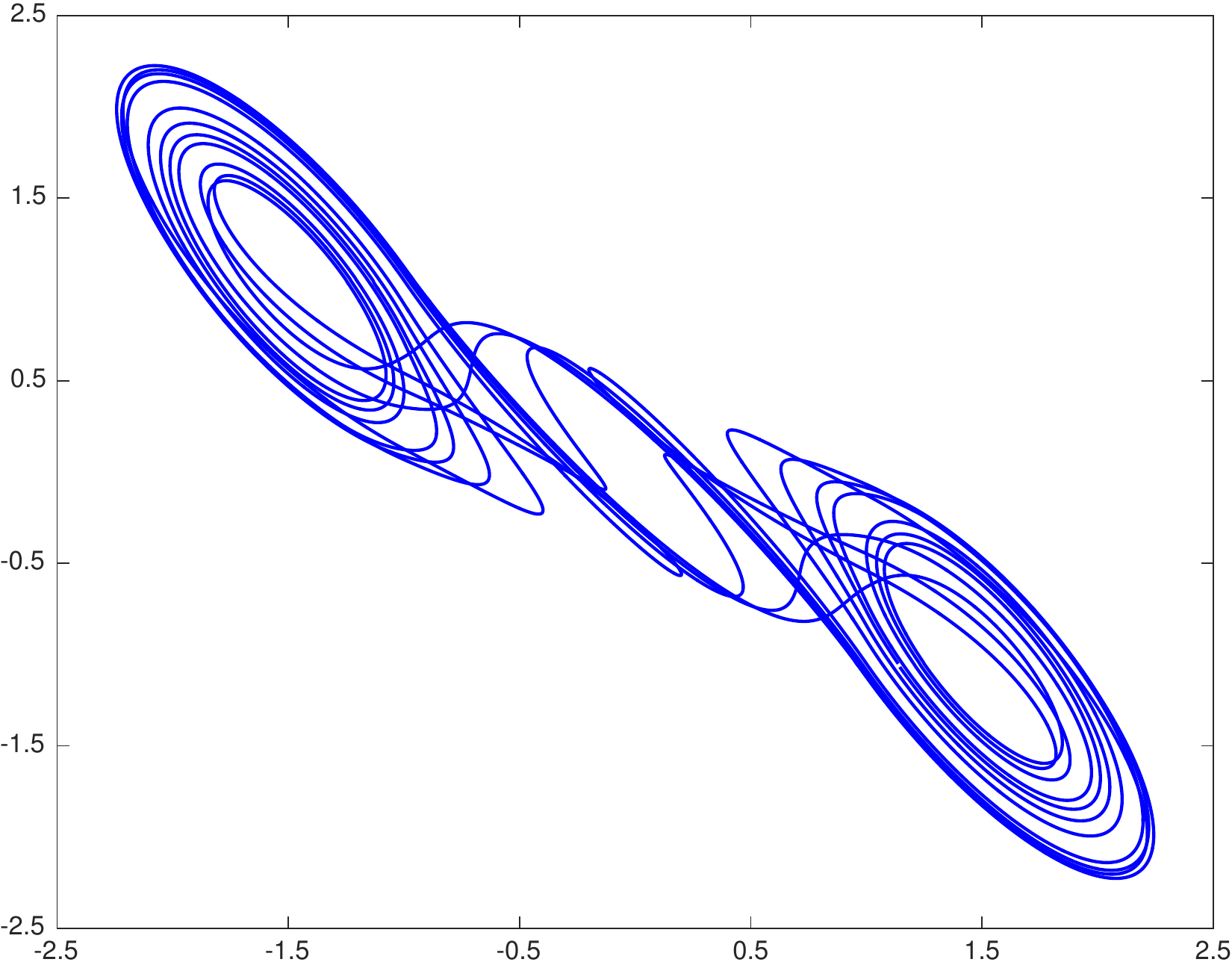} &
		\includegraphics[width=0.425\textwidth]{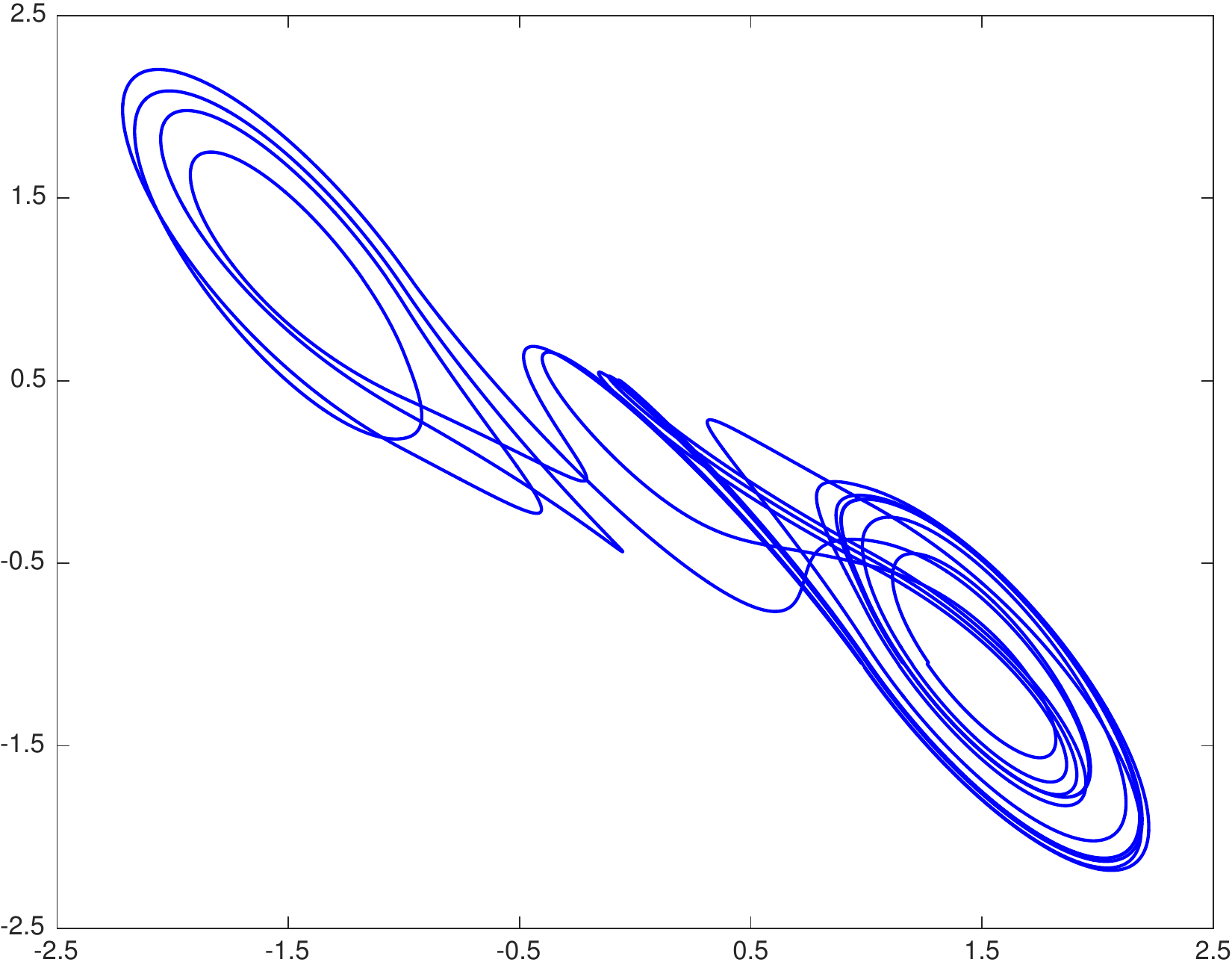} \\
		(a) & (b) \\
	\end{tabular} 
	\caption{Entangled cupolets (a) $\mathbf{C}00000000011$ and (b) $\mathbf{C}0000110011110011$ with visitation sequences
	$\mathbf{V}0000011100011111000111$ and $\mathbf{V}0000111111111111$, respectively~\cite{Morena2014b}.{\label{fig:example_C00000000011-C0000110011110011}}}
\end{figure*}

\begin{figure*}[!t]
	\centering
	\small
	\begin{tabular}{cc}
		\frame{\includegraphics[width=0.475\textwidth]{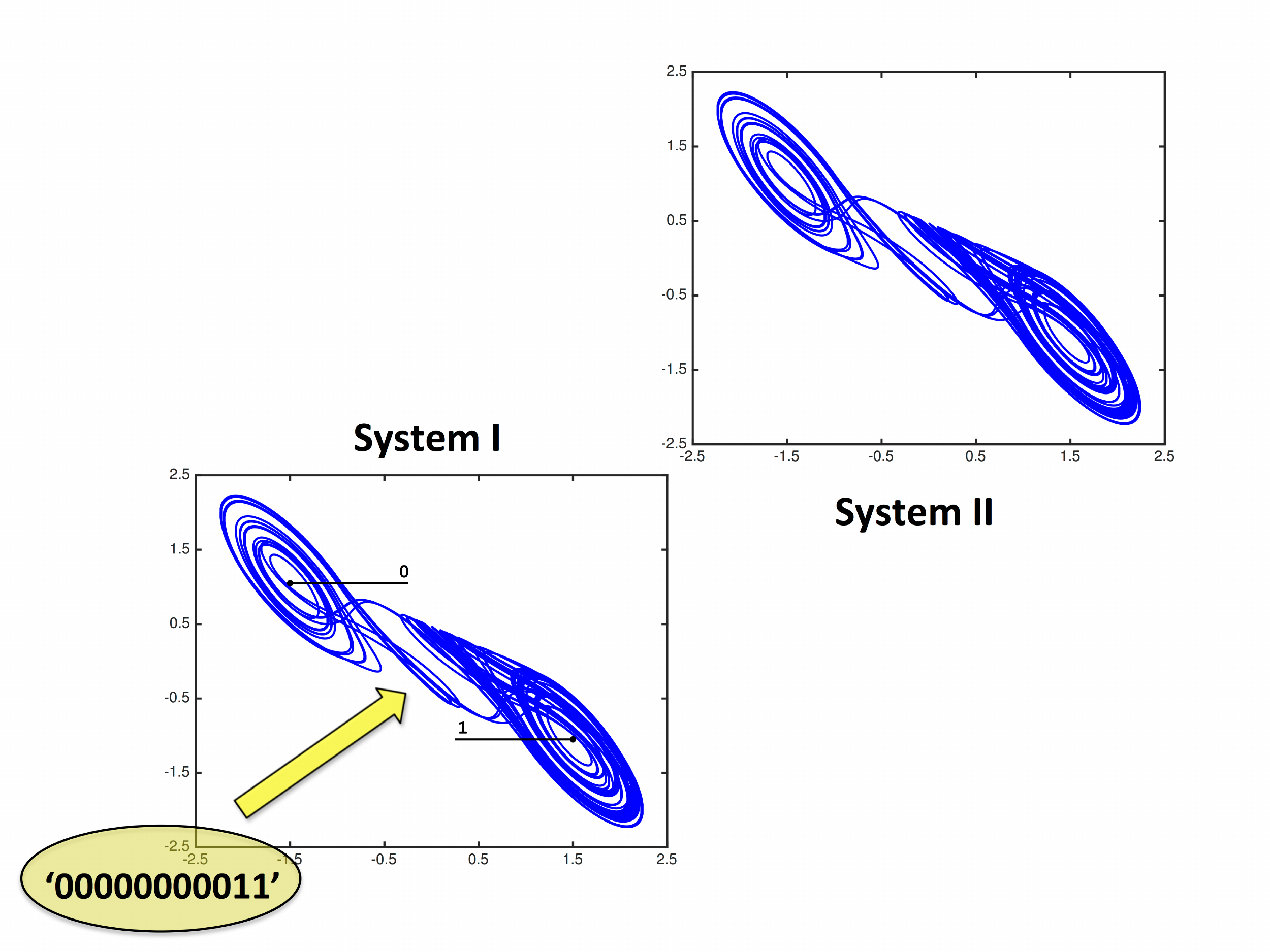}} &
		\frame{\includegraphics[width=0.475\textwidth]{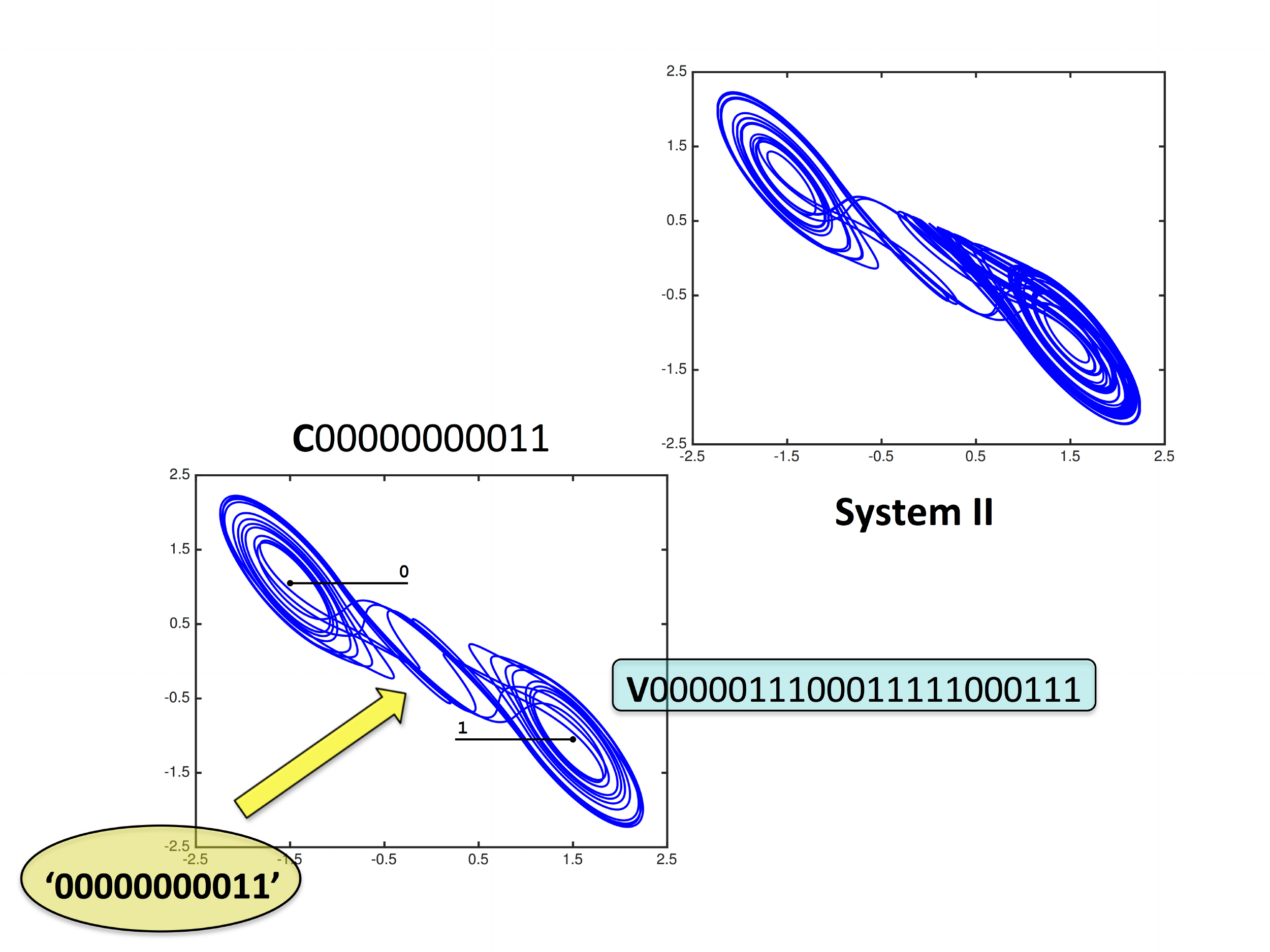}} \\
		(a) & (b) \\[2ex]
		\frame{\includegraphics[width=0.475\textwidth]{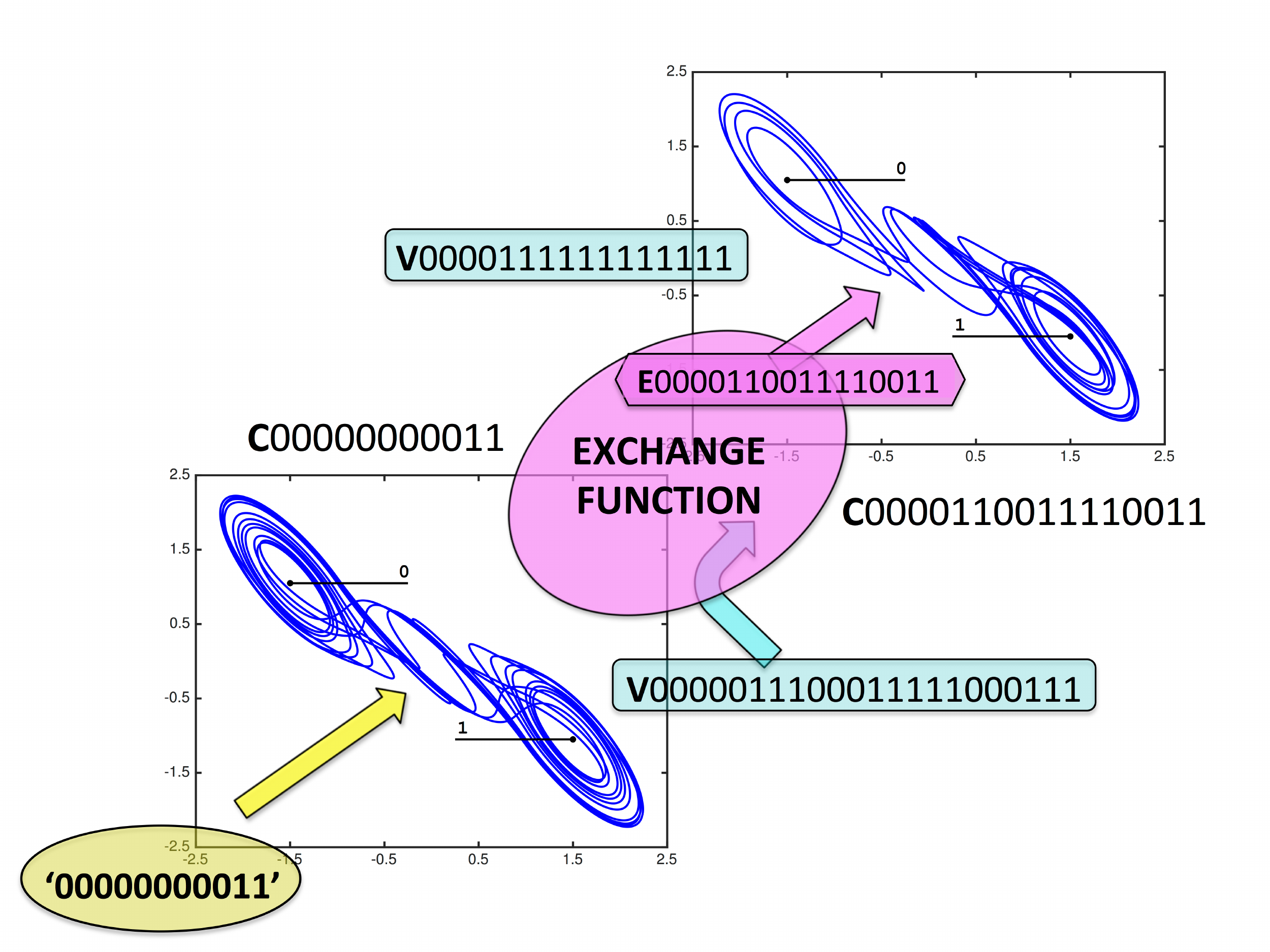}} &
		\frame{\includegraphics[width=0.475\textwidth]{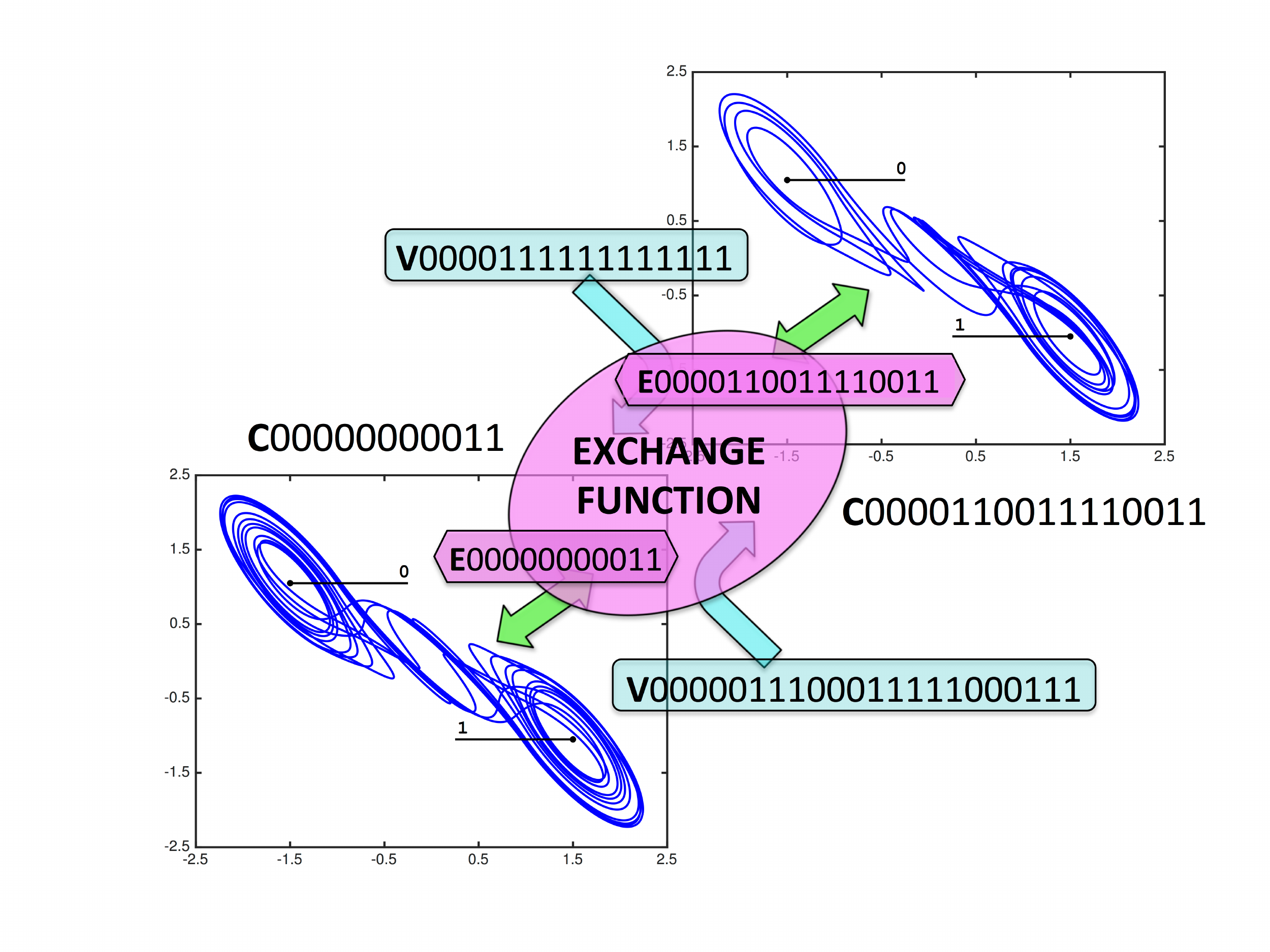}} \\
		(c) & (d)
	\end{tabular}	
	\caption{(Color online) Schematic illustration of chaotic entanglement: in (a) a control sequence is externally applied to a chaotic system,
System~I, via the indicated (yellow) control pump. System~I subsequently stabilizes in (b) onto cupolet $\mathbf{C}00000000011$ according to the
control method described in Section~\ref{sec:cupolet_background}. This cupolet then evolves around the attractor to generate its visitation sequence,
$\mathbf{V}0000011100011111000111$. In (c), an exchange function accepts this visitation sequence as an input and the outputted emitted sequence,
$\mathbf{E}0000110011110011$, is taken as a control sequence and used to control a second chaotic system, System~II. This subsequently induces
System~II to stabilize uniquely onto cupolet $\mathbf{C}0000110011110011$ whose visitation sequence, $\mathbf{V}0000111111111111$, is then passed to
the same exchange function in (d). The resulting emitted sequence, $\mathbf{E}00000000011$, is applied as control instructions to
$\mathbf{C}00000000011$ of System~I. Each emitted sequence exactly matches each corresponding cupolet's control sequence, and so the external control
pump seen in (a) is unnecessary and can be removed. Systems~I and~II are now dynamically engaged in a state of perpetual mutual-stabilization between
their respective cupolets and are thus considered chaotically entangled. This entanglement is summarized in
Table~\ref{tab:chaotic-entanglement}.{\label{fig:chaotic-entanglement}}}
\end{figure*}

As an example, we will demonstrate how the two cupolets shown in Figure~\ref{fig:example_C00000000011-C0000110011110011} can become chaotically
entangled. This process is also depicted in Figure~\ref{fig:chaotic-entanglement} as a series of step-by-step illustrations. First, two double scroll
systems, Systems~I and~II, are simulated without control. In order to stabilize one of these cupolets, say $\mathbf{C}00000000011$, the control
sequence `00000000011' must be applied to System~I using the cupolet-generating control technique described in Section~\ref{sec:cupolet_background}.
This step is illustrated in Figure~\ref{fig:chaotic-entanglement}(a), where the depiction of the control planes indicates that System~I is being
controlled via the (yellow) external control pump. Once $\mathbf{C}00000000011$ completes one full period around the attractor, its visitation
sequence, $\mathbf{V}0000011100011111000111$, is realized. Figure~\ref{fig:chaotic-entanglement}(b) captures this stage of the entanglement process.
This visitation sequence is then passed to an exchange function where it is modified according to a predefined binary operation and sent to System~II
as an emitted sequence. In this example, a `complement' exchange function converts $\mathbf{V}0000011100011111000111$ into the emitted sequence
$\mathbf{E}0000110011110011$ essentially by interchanging subsequences of ones and zeros for zeros and ones, respectively.\footnote{This particular
type of exchange function is more thoroughly described in~\cite{Morena2014b}.} As the emitted sequence passes from the exchange function, it is
applied to System~II as instructions for controlling this system. In this case, System~II stabilizes onto the cupolet $\mathbf{C}0000110011110011$
since the emitted sequence actually is this second cupolet's control sequence. These particular steps are visualized in
Figure~\ref{fig:chaotic-entanglement}(c).

The cupolets' interaction now repeats in the reverse direction. The visitation sequence of cupolet $\mathbf{C}0000110011110011$ is found to be
$\mathbf{V}0000111111111111$ (see Figure~\ref{fig:chaotic-entanglement}(c)), which is converted by the same exchange function into the emitted
sequence $\mathbf{E}00000000011$. When applied as a control sequence to System~I, $\mathbf{E}00000000011$ preserves the stability of cupolet
$\mathbf{C}00000000011$ because the emitted bits, `00000000011', match the control information needed to maintain this cupolet's stability. This
two-way exchange of control information between Systems~I and~II defines the cupolets' interaction which has been managed by the exchange function.
Notice that both emitted sequences, $\mathbf{E}00000000011$ and $\mathbf{E}0000110011110011$, match the required control sequences for cupolets
$\mathbf{C}00000000011$ and $\mathbf{C}0000110011110011$, respectively. The (yellow) external control pump is thus rendered redundant and can be
discarded now that the cupolets are dynamically generating the necessary control instructions themselves. Since the cupolets are effectively driving
each other's stability, they are considered chaotically entangled and their stabilities are guaranteed so long as their two-way interaction is
undisturbed. Figure~\ref{fig:chaotic-entanglement}(d) illustrates this final step of the entanglement and Table~\ref{tab:chaotic-entanglement}
summarizes the correspondence between the control, visitation, and emitted sequences of each cupolet.

\begin{table}[!t]
	\centering
	\caption{(Color online) The following table summarizes the chaotic entanglement induced between two interacting cupolets, $\mathbf{C}00000000011$
(of System~I) and $\mathbf{C}0000110011110011$ (of System~II). The orbits of these cupolets are depicted in
Figure~\ref{fig:example_C00000000011-C0000110011110011}, while the generation of the entanglement via a `complement' exchange function is illustrated
in Figure~\ref{fig:chaotic-entanglement}. Notice that the control sequence required to sustain the stability of cupolet $\mathbf{C}00000000011$ is
contributed by cupolet $\mathbf{C}0000110011110011$ via this cupolet's emitted sequence, $\mathbf{E}00000000011$. Similarly, the stability of
$\mathbf{C}0000110011110011$ is maintained by the repeated application of emitted sequence $\mathbf{E}0000110011110011$, which is generated by
$\mathbf{C}00000000011$ via the same exchange function. The font colors in this table are intended to accentuate the correspondence between the
cupolets' control sequences and their emitted sequences. Details of the entanglement generation are found in the text.\vspace*{1ex}
{\label{tab:chaotic-entanglement}}}

	\begin{tabular}{@{}llll@{}}
		\br
		 & \textbf{Cupolet} & \textbf{Visitation Sequence} & \textbf{Emitted Sequence} \\
		\mr
		\textbf{System~I} & $\mathbf{C}\textcolor{blue}{00000000011}$ & $\mathbf{V}0000011100011111000111$ & $\mathbf{E}\textcolor{red}{0000110011110011}$ \\
		\textbf{System~II} & $\mathbf{C}\textcolor{red}{0000110011110011}$ & $\mathbf{V}0000111111111111$ & $\mathbf{E}\textcolor{blue}{00000000011}$ \\
		\br
	\end{tabular}
\end{table}

Strictly speaking, chaotic entanglement need not be associated exclusively with interacting cupolets because its generation extends naturally to pairs
of interacting UPOs. Cupolets represent highly accurate approximations to these UPOs, but in general two interacting chaotic systems chaotically
entangle once each system stablizes onto a particular UPO whose stability is then maintainted by the symbolic dynamics of the partner UPO. The
visitation sequences of the UPOs would continue to provide an appropriate symbolic dynamics, but the advantage of inducing and detecting entanglement
with cupolets is twofold.

First, the control technique described in Section~\ref{sec:cupolet_background} is designed to stabilize cupolets. In doing so, the technique makes
accessible the symbolic dynamics of chaotic systems while greatly simplifying how the interactions between the systems are simulated. For instance,
perturbations are applied only when a cupolet intersects a control plane, which means that the cupolet's remaining evolution is freely determined by
the system's governing equations. Therefore, when detecting entanglement between two systems, one only needs to monitor finitely-many intersections
with the control planes, which allows one to simultaneously observe the visitation sequence of each evolving cupolet. Second, given that cupolets can
be generated very efficiently, a great deal of useful information can be collected simply by recording the control and visitation sequences of a
sufficiently large collection of pre-generated cupolets. This has been shown to facilitate the detection of candidate cupolets with which a given
cupolet could entangle~\cite{Morena2014b}.


\subsection{Chaotic entanglement as an analog of quantum entanglement}
\label{sub:chaotic_entanglement_as_an_analog_of_quantum_entanglement}

Chaotic entanglement exhibits many properties that are characteristic of quantum entanglement~\cite{Peres1993,Nielsen2000,Haake2001}. For instance,
measurements that disrupt the interaction between two entangled cupolets, say $\textbf{C}_{\rm{A}}$ and $\textbf{C}_{\rm{B}}$, will almost always
destroy their entanglement unless a great deal is known about the control scheme. By measurement, we mean a perturbation that could be as meticulous
as the macrocontrols or microcontrols integral to the control scheme described in Section~\ref{sec:cupolet_background}, or as general as an arbitrary
perturbation applied to the system. As an example, consider the subtle effect of interchanging a `0' bit for a `1' bit in the control sequence of
$\textbf{C}_{\rm{A}}$, or vice versa. Control sequences are unique since they direct a chaotic system onto one specific cupolet. Therefore, disturbing
the cupolet's control sequence would perturb its trajectory into a different bin on the control plane and cause $\textbf{C}_{\rm{A}}$ to either
destabilize or to transition to a different cupolet as described in Section~\ref{sub:application_of_cupolets}. In either scenario,
$\textbf{C}_{\rm{A}}$ produces a different visitation sequence that no longer guarantees the stability of the partner cupolet $\textbf{C}_{\rm{B}}$,
and so the entangled state is lost. However, should the appropriate controls for cupolet $\textbf{C}_{\rm A}$ be restored and continue to be
periodically applied, then $\textbf{C}_{\rm A}$ will eventually restabilize followed by $\textbf{C}_{\rm B}$ via the exchange function.

In Section~\ref{sec:main_discussion}, we describe how a measurement can be carefully designed so that its effects do not significantly disturb the
stability of the intended cupolet. Doing so would allow one to probe a pair of entangled cupolets without compromising their entanglement and would
also provide access to the control sequences that are uniquely associated to each cupolet. In other words, having full knowledge of the control
mechanism would permit the control information stored in an entangled state to be recovered and read later. In this way, entangled cupolets remember
the state of the control bits that are originally used in establishing their entanglement and naturally form a memory device for information. This
process of inserting, storing, and retrieving information in pairs of entangled cupolets is consistent to what is currently being developed with
quantum computing~\cite{Nielsen2000,Pittman2002}.



\subsection{Pure chaotic entanglement}
\label{sub:pure_entanglement}

In some instances, chaotic entanglement occurs without the assistance of an exchange function (or, equivalently, via an identity exchange function).
This is known as \emph{pure entanglement} because it requires no environmental property in order to be induced or
sustained~\cite{Morena2014b,Morena2014c}. Instead, a visitation sequence is converted directly to an emitted sequence without any intermediary
modification being made. Since every visitation sequence simply represents a bit string indicating which lobe is being visited, we assume that any time
the $1$-lobe of cupolet $\textbf{C}_{\rm A}$ is visited, energy accumulates and a perturbation is applied directly to cupolet $\textbf{C}_{\rm B}$ (and
vice versa). That is, each purely-entangled cupolet generates the exact sequence of control bits necessary for maintaining its partner's periodic orbit
without any assistance from an exchange function, but simply by realizing its own visitation sequence. This makes pure entanglement the simplest form
of cupolet entanglement. Entanglement induced with the aid of an exchange function is considered a variation of pure entanglement because a nontrivial
operation must be performed on a cupolet's visitation sequence in order to generate an emitted sequence.

Existence of pure chaotic entanglement has been documented in~\cite{Morena2014b,Morena2014c} and indicates the potential for such behavior to arise
naturally and spontaneously between interacting chaotic physical systems. This may not be altogether surprising given that experimental evidence of
natural and macroscopic quantum entanglement has recently been reported in~\cite{Arnesen2001,Vedral2004,Vedral2008}. With no external controller or
exchange function needed as a catalyst, it is possible that such direct cupolet-to-cupolet interactions may arise spontaneously and lead to naturally
entangled states. Spontaneous chaotic entanglement is further discussed in Section~\ref{sec:main_discussion} where we consider additional connections
between chaotic and quantum systems.




\section{Main discussion: quantum parallels in chaotic systems}
\label{sec:main_discussion}

While chaotic entanglement demonstrates a new way for chaotic systems to interact, this behavior also signals a new parallel between quantum and
classical mechanics. In the following discussion, we explore other such connections and emphasize that chaotic systems are capable of supporting
nonclassical behavior. We also address several concerns which invariably arise when examining quantum systems for entanglement. Key to our discussion
is the important role that cupolets and, by extension, UPOs of chaotic systems play in determining the properties and dynamics of chaotic systems.

\subsection{Hilbert space considerations}
\label{sub:hilbert_space_considerations}

Formulating a Hilbert space of states is taken as a starting point in many quantum studies because it is from these vector spaces that an associated
wave function can be expressed as a linear combination of state vectors. A typical way to express these vector spaces is via the Fourier modes of the
system or by postulating linear combinations of sinusoids in order to define the state vectors. For chaotic systems, the only meaningful states are
solutions of their governing equations. One could define the cupolets to be the state vectors, or more precisely close approximations to the state
vectors, except that then the superposition of state vectors could not be guaranteed due to the system's inherent nolinearity. Since there are many
thousands of cupolets that can be isolated, these orbits would also appear to form an overdetermined set of basis elements. As
Figure~\ref{fig:cupolet_diversity} illustrates, the Fourier spectra obtained from any large collection of cupolets would be extremely diverse: the
simplest cupolets exhibit one or two significant spectral peaks, while more complex cupolets have tens or hundreds of significant peaks in their
spectra. Although it would seem that Fourier modes and the related Hilbert space ideas should hold for chaotic systems, the reality is that cupolets
are collectively overdetermined and do not satisfy any simple orthogonality principle.


\newpage

\subsection{Superposition of states}
\label{sub:superposition_of_states}

Rather than considering cupolets as the actual state vectors, a more natural way to describe the wave function of a chaotic system is by considering
the cupolets as the ``states'' of the system and then by formulating vectors of cupolet states. This is reasonable because chaotic systems typically
admit a dense, countably infinite set of UPOs on their attractors, and cupolets represent highly-accurate approximations to the UPOs. Also, ergodicity
guarantees that a free-running chaotic system eventually realizes all possible non-equilibrium states and visits arbitrarily small neighborhoods of
its periodic solutions. Even though chaotic systems evolve aperiodically for all time, the dynamics of these systems are ultimately confined to their
attractors, which means that a wandering chaotic trajectory undergoes a series of close encounters with the UPOs and cupolets.

Collectively, UPOs provide a rich source of qualitative information about their associated chaotic system. Many characteristics of chaotic systems,
such as Lyapunov exponents, the natural measure, dimension, topological entropy, and several orbit expansions, can all be expressed in terms of these
periodic orbits~\cite{Eckmann1985,Auerbach1987,Grebogi1988,Ornstein1991,Ruelle1999,Franzosi2005}. In particular, the natural measure, which can be
loosely interpreted as the probability of a chaotic system visiting a given region of its attactor over time, is often described as being concentrated
on the UPOs. In other words, as it evolves, a chaotic system visits regions populated by UPOs with greater frequency and will dwell alongside an UPO
for an extended amount of time after which the trajectory begins shadowing other UPOs. Furthermore, since UPOs are solutions to the governing
differential equations, uniqueness properties imply that these orbits cannot be crossed in phase space. This is easier to visualize for many
low-dimensional chaotic systems, such as the double scroll, Lorenz, and R\"{o}ssler systems, where the associated attractor is locally ribbon-like in
at least part of its domain. Cupolets are generated in such a way that uniqueness considerations also apply, except possibly at certain locations
along a control plane where the controls are applied.\footnote{See~\cite{Morena2014a} for more technical details behind cupolet intersections.}
Therefore, chaotic trajectories are restricted to evolving along unique paths that are locally bounded by both UPOs and cupolets, and so one may
consider the dynamics of chaotic systems to be linearly dependent on these orbits.

To represent the solution or \emph{state} of a given chaotic system as a vector of cupolet states, let $\varphi_{k}=\varphi_{k}(t)$ denote the state of
a particular cupolet, $\textbf{C}_{k}$, and let $\Phi=\Phi(t)$ denote the state of the chaotic system, both at time $t\in \mathbb{R}$. Here, $k\in
\mathbb{N}$, although a measure zero set of periodic solutions is not always admitted due to restrictions inherent in a chaotic system. Taking
$\varphi_{k}$ and $\Phi$ to represent the state space coordinates of the cupolet and a chaotic trajectory respectively, the state of a chaotic system
may be formulated as a weighted sum of its cupolets:
\begin{equation}
	\displaystyle
	\mbox{\large $\Phi$} = \sum_{k=1}^{\infty} \alpha_{k}\varphi_{k},
	\label{eq:cupolet_expansion}
\end{equation}
where each weight, $\alpha_{k} \in \mathbb{R}$, represents the contribution to $\Phi$ from cupolet $\textbf{C}_{k}$ at time $t$. As the chaotic system
evolves in time, each $\alpha_{k}$ fluctuates depending on the proximity of the system to each cupolet. One can never exactly ascertain the state of a
chaotic system since chaos effectively represents a mixture of a dense set of UPOs, which means that the state of a chaotic system is best described as
suspended in an evolving \emph{superposition} of its periodic orbits. For instance, when the system is dwelling near cupolet $\textbf{C}_{k}$, then
$\Phi \approx \varphi_{k}$ because at this moment $\alpha_{k} \neq 0$ and $\alpha_{l}\approx 0$ for all $l \neq k$. A chaotic system's \emph{state
vector}, $\vec{\alpha} \in \mathbb{R}^{\infty}$, may now be constructed by considering the weights of each cupolet:
\begin{equation}
	\vec{\alpha} = \left(\alpha_{1},\alpha_{2},\ldots, \alpha_{k},\ldots \right).
	\label{eq:cupolet_vector}
\end{equation}
This vector provides a complete and evolving description for the state of a chaotic system in terms of its cupolets (or equivalently, its UPOs). In
this way, an uncontrolled, or freely evolving, chaotic system may be viewed as evolving in a ``mixed state'' that is a linear combination or
superposition of cupolet states.

In quantum mechanics, the wave function is of fundamental importance since it provides a probabilistic description of the state of a quantum system.
The analog for a chaotic system is its state vector, $\vec{\alpha}$. The UPOs and cupolets of chaotic systems thus capture important information about
their parent chaotic system, while also providing a deterministic description for the state of the system, as evidenced by
Equation~(\ref{eq:cupolet_expansion}).

\subsection{Wave function collapse}
\label{sub:wave_function_collapse}

Another important concern in quantum mechanics is the idea that making a measurement causes the collapse of a quantum system's associated wave function
onto a specific state. Prior to the disturbance, the wave function is suspended in a superposition of state vectors, meaning that the quantum system
cannot be unambiguously described. Analogously, the collapse of a chaotic system onto a particular state would occur exactly when the system stabilizes
onto a cupolet, say $\textbf{C}_{k}$. Via Equation~(\ref{eq:cupolet_expansion}), when this happens, $\alpha_{k}=1$ and $\alpha_{l}=0$ for all $l \neq
k$, which gives $\Phi=\varphi_{k}$ as desired. The state vector given by Equation~(\ref{eq:cupolet_vector}) reduces as well to
$\vec{\alpha}=(0,\ldots,0,1,0,\ldots)$, whose only nonzero component is its $k^{th}$. Until this collapse occurs, the chaotic system cannot be
definitively described as a single cupolet state. This parallel notion of wave function collapse in chaotic systems is further supported by the
following key observations.

First, cupolets represent approximate periodic orbits of the system, and while these periodic orbits are dynamical solutions of the defining
equations, they are all unstable. Second, when controls are applied to a chaotic system, cupolets arise because of two unexpected properties: the
system stabilizes uniquely onto a cupolet under the influence of a set of repeating perturbations, and this stabilization occurs independently of
initial conditions. These properties allow a chaotic system to be collapsed onto a specific cupolet from any initial point. Third, consider that a
measurement process applied to any system that exists at the scale where either chaotic or quantum effects may be observed is likely to perturb the
system in some prescribed fashion. We assert that this sort of measurement would have an effect similar to the control process that is used for
stabilizing cupolets. By viewing a chaotic system as a superposition of cupolet states, repeated applications of these measurements would lead to the
collapse of the system's state vector onto a particular cupolet state. This would occur precisely when the chaotic system stabilizes uniquely onto a
cupolet, thereby evoking a parallel to the collapse of a quantum wave function.


\subsection{Natural and spontaneous entanglement}
\label{sub:natural_and_spontaneous_entanglement}

The concepts of measurement and state vector collapse need not be induced by external measurements or by user-implemented controls, but could well
occur naturally in chaotic entanglement. Although isolated chaotic systems evolve aperiodically because their periodic orbits are unstable, a chaotic
system tends to dwell significantly longer on its UPOs than on its other states or regions of phase space. By extension, an ensemble of independent
chaotic systems would also be dwelling along their UPOs and cupolets infinitely often. If one chaotic system happens to dwell on a cupolet that not
only exhibits the ability to entangle, but that can also communicate control information to a second nearby chaotic system, and if this interaction is
as successful in the reverse direction, then these two systems would entangle naturally.

In the context of two arbitrary cupolets, $\textbf{C}_{\rm A}$ and $\textbf{C}_{\rm B}$, this situation implies that the parent system of cupolet
$\textbf{C}_{\rm A}$ will approach and dwell on $\textbf{C}_{\rm A}$ infinitely often. If a second chaotic system is at the same time dwelling near
$\textbf{C}_{\rm B}$, then entanglement could form spontaneously between the two systems, provided that the symbolic dynamics of the cupolets can be
used to maintain their periodic behavior. In this way, isolated and independently-evolving chaotic systems would be perturbing each other with the
interactions themselves playing the role of the controls or measurements. This makes it possible for entanglement to occur spontaneously, as has been
emphasized both in Section~\ref{sub:pure_entanglement} and in the recent studies of macroscopic systems examined
in~\cite{Arnesen2001,Vedral2004,Vedral2008}. As we discuss below, the potential for spontaneous chaotic entanglement plays a key role in the
interpretation of making measurements on individual members of entangled cupolet pairs.


\subsection{Measurement problem}
\label{sub:measurement_problem}

It is first worthwhile to compare the effects of a \emph{knowledgeable measurement} on a chaotic system, as opposed to a \emph{blind measurement}. For
instance, if one has both knowledge of this control scheme and access to measurement tools that are smaller than the scale of the control bins, then
one could monitor the state of the chaotic system without disturbing its trajectory. A particular measurement could be carefully designed so that its
effects would not be strong enough to perturb an evolving cupolet to a new bin center on a control plane; any slight deviation from the original orbit
would be corrected the next time the cupolet intersects a control plane via the implementation of the microcontrols. This is known as a knowledgeable
measurement as it permits one to study a cupolet without compromising its stability and thus also allows one to probe two entangled systems and not
compromise the entanglement.

Were a measurement not implemented as carefully, the repercussions may be much more pronounced. For example, consider the effects of the measurement
previously described in Section~\ref{sub:chaotic_entanglement_as_an_analog_of_quantum_entanglement}, whereby a single `1' control bit in a given
cupolet's control sequence is changed to a `0' control bit. Such a disturbance would destabilize the cupolet and cause the parent system to either
revert to chaotic behavior or to stabilize again after a potentially long transient period. This disturbance is known as a blind measurement and it
would cause the destablized orbit to begin realizing a new visitation sequence. If this cupolet had been entangled with another cupolet, then the
effects of the blind measurement would transfer to the partner cupolet by way of the exchange function, which would begin producing a different
emitted sequence. The unfamiliar emitted sequence would no longer match the control sequence required to ensure the stability of the partner cupolet,
and so the entanglement would be lost.

Regarding the measurement problem, consider the case of a pair of entangled cupolets, where the cupolets have entangled either through the user
preparation of an entangled state or naturally through pure entanglement. If a knowledgeable measurement is conducted on one member of the entangled
pair, then the state of the other member would be known with certainty (with the proviso that we have only found unique pairings at this point). If
the measurement process involves blind measurements, however, then the disrupted communication between the members of the entangled pair would induce
the two parent systems to begin evolving independently. Similarly, if one postulates that the interaction between members of an entangled pair is
limited by distance, and if the entangled cupolets are separated too far, the systems would drift away from their state of entanglement as their
communication wanes. This drift would not necessarily be very rapid, but would be determined by the local Lyapunov exponents of the two
cupolets~\cite{Franzosi2005}. In this situation, the history of the two chaotic systems' previous entanglement would not be immediately erased because
a measurement conducted on one member of the entangled pair would be predictive of the state of the second system, although the accuracy of the
prediction would decay over time.

In contrast, the principles of quantum mechanics dictate that making any measurement on a system immediately alters its state. This is problematic for
researchers when determining the actual state of a quantum system~\cite{Peres1988,Chefles2000}. As indicated by Isham,
\begin{quote}
	``$\ldots$ quantum theory encounters questions that need to be answered, one of the most important of which is what it means to say, and how it can be ensured, that the individual systems on which the repeated measurements are to be made are all in the `same' state immediately before the measurement.  This crucial problem of \emph{state preparation} is closely related to the idea of a reduction of the state vector.''~\cite{Isham1995}
\end{quote}
When combined with knowledgeable measurements, the cupolet-stabilizing control scheme could specifically help in state preparation for experiments.
Cupolets can be generated regardless of the current state of the system, which means that if chaotic control methods are designed to stabilize
cupolets from physical systems, then it would be possible to prepare systems in the same state prior to experimental measurements. Experimenters could
then probe further into the classical--quantum transition without interrupting an entanglement state.


\subsection{Entropy}
\label{sub:entropy}

The last feature of quantum systems that we would like to examine from a classical perspective is entropy. In quantum mechanics, entropy is often used
to assess entanglement strength~\cite{Chaudhury2009}, yet our discussion has so far focused primarily on the actual states of chaotic systems, rather
than on quantifying a chaotic entanglement. In chaotic systems, one may define entropy according to the methods of
Kolmogorov~\cite{Eckmann1985,Ornstein1991,Short1993}. Kolmogorov entropy quantifies the rate of growth of information as a dynamical system evolves
over time. Such information is typically encoded in the system's symbolic dynamics, which means that this type of entropy is related to the growth
rate of symbolic sequences generated by a chaotic system. For the double scroll system, the visitation sequences provide such a symbolic dynamics.

A periodic orbit, for example, would be assigned zero entropy because even though it would repeat forever, its rate of growth of information becomes
zero once the lengths of its symbolic sequences exceed the orbit's period. For truly random systems, this rate of growth goes to infinity in the sense
that all possible sequences are realized as the system evolves. In between are the chaotic trajectories, whose dynamics never repeat, yet the geometry
of chaotic attractors is such that the rate of growth of the symbolic sequences is finite and thus so is the entropy. Accordingly, a chaotic system in
its natural, uncontrolled, and isolated state generates entropy at a finite rate, but once it is controlled by the cupolet-generating control scheme,
or if it enters into an entangled state, the system immediately collapses onto a particular cupolet or UPO and has zero entropy from that point
onward~\cite{Hunt2015}. This demonstrates that chaotic entanglement does admit notions of entropy and is effectively an entropy-reversing operation.


\subsection{Differences with quantum entanglement}
\label{sub:differences_with_quantum_entanglement}

Now that we have discussed several properties of chaotic systems that draw parallels between chaotic and quantum entanglement, it is important to note
that there are a number of differences as well. First, superposition in a purely quantum sense refers to linear combinations of state vectors that
collectively describe the state of a quantum system and that satisfy the Schr\"{o}dinger equation. In chaotic systems, conventional superposition is
not permitted because the governing equations are inherently nonlinear. Superposition instead refers to a chaotic system existing as a mixture of its
cupolets, or more precisely, its UPOs. In this framework, the state of a chaotic system can be well-represented as a linear combination of the states
of these periodic orbits. As the chaotic system evolves in time, so too does its state vector since each $\alpha_{k}$ in
Equation~(\ref{eq:cupolet_vector}) represents the contribution of an associated cupolet or UPO to the current state of the chaotic system.

A second difference between quantum and chaotic entanglement concerns how expectedly entanglement can arise. Quantum entanglement is typically created
deliberately between subatomic particles via direct interactions like atomic cascades or spontaneous parametric down-conversions~\cite{Clauser1978,
Horodecki2009}. For chaotic entanglement to arise, interaction is required because two chaotic systems must be able to communicate control information
to each another. One could deliberately arrange two interacting chaotic systems into entanglement by implementing the chaotic control scheme described
in Section~\ref{sec:cupolet_background} and then by monitoring the ensuing interaction. Chaotic entanglement may also form spontaneously and without
the aid of any external preparation or control. As discussed in Section~\ref{sub:pure_entanglement}, this is known as pure entanglement, and it arises
because evolving chaotic systems are constantly visiting neighborhoods of their periodic orbits. This increases the likelihood that two interacting
chaotic systems may be concurrently shadowing periodic orbits that could become mutually-stabilizing. If so, then the two systems could collapse into
entanglement without any external preparation required.

Unlike quantum entanglement, which allows for nonlocal correlations, chaotic entanglement is neither distance-independent nor instantaneous in its
response to measurements. Although the chaotic entanglement that we have documented in~\cite{Morena2013,Morena2014b} exists strictly at the
information-theoretic stage, we have discussed its potential to be detected in physical systems. As such, chaotic systems must necessarily be evolving
in close proximity in order for their communication to induce an entanglement. Should two chaotically-entangled systems become physically separated,
or simply lose the ability to communicate, then the efficacy of their interaction diminishes to zero, leading to a loss of entanglement. Each system
would then revert to chaotic, aperiodic behavior because their trajectories are no longer being directed along the periodic orbits on which the
systems had been previously stabilized. Despite the communication breakdown, the deviation from the periodic orbits will not be instantaneous, but
gradual. A chaotic trajectory naturally tends to dwell alongside nearby UPOs, which means that after the entanglement has been interrupted, each
previously-entangled system will continue evolving in close proximity of its UPO for a period of time proportional to the Local Lyapunov Exponent of
that periodic orbit. One would therefore not expect chaotic entanglement to exhibit instantaneous action at a distance.

This delayed response to measurements is additionally interesting because it allows a disrupted chaotic entanglement to be reacquired between two
previously-entangled chaotic systems. If the interaction is restored between the two systems, either by shortening their spatial separation or by
removing any communication barriers, then the two systems may not necessarily have drifted too far from their previously-stabilized periodic orbits,
especially if their communication is restored quickly enough. With their interaction reinstated, the two systems could redirect each other back onto
their respective periodic orbits, thus reinstating the entanglement.

In summary, the key ingredients missing from our list of classical analogs to quantum mechanical characteristics are instantaneous responses to
measurement and nonlocality. In quantum mechanics, when a measurement is applied to one of two entangled particles, the state vectors of both
particles each instantly collapse onto a specific state vector regardless of spatial separation. In contrast, chaotic entanglement is limited by
physical distances and exhibits a delayed response to disturbances due to the influence UPOs have on the dynamics of chaotic systems.



\section{Concluding remarks}
\label{sec:concluding_remarks}

For several decades, it has been a goal of many mathematicians and physicists to establish connections between classical chaos and quantum physics.
Some researchers have conjectured that nonlinearities may help explain the paradoxes of certain Bell inequalities that arise in quantum
mechanics~\cite{McHarris2003,McHarris2011}, while other researchers have detected chaotic behavior in true quantum settings~\cite{Chaudhury2009}. The
research that we recently documented in~\cite{Morena2014b} and that we have discussed in this paper considers the classical--quantum correspondence
from a classical perspective.

Though cupolets are themselves periodic orbits which have been stabilized from a chaotic system, the parallels that chaotic entanglement evokes with
quantum entanglement are worthy of consideration. For example, any measurement not possessing full knowledge of the cupolet control scheme would
destroy the entanglement, yet detailed knowledge would allow the control information stored in entangled cupolets to be recovered without compromising
the entanglement. Furthermore, although cupolets could not be used to rigorously formulate a conventional Hilbert space model for an associated chaotic
system, the state vectors of chaotic systems could be represented as superpositions of cupolets. In this framework, the quantum notions of measurement,
entanglement, and collapse of a wave function are all relevant to chaotic systems, as is entropy, a standard way of measuring entanglement.

This identification of quantum signatures in chaotic systems can be pushed a long way, only to reach a limit when nonlocality is considered. In order
to detect chaotic entanglement in interacting physical systems, the interaction cannot be spatially separated beyond a communication horizon, nor can
one expect the entanglement to arise instantaneously. As a result, it seems unlikely that a classical analog of nonlocality may ever be established.
Even so, there is merit in examining quantum mechanics from a classical perspective. Doing so allows one to identify the features of entanglement that
are quintessentially quantum mechanical and to appreciate the unique role that nonlocality plays in quantum mechanics. There may be other discrepancies
as well, and it is hoped that these differences may be used to detect whether an observed entanglement is produced by a quantum process, or whether
there may be an underlying chaotic process at work.

The key point is that a classical version of entanglement has been observed from among the dense set of UPOs of a typical chaotic system. This is
significant because the properties of chaotic behavior actually increase the likelihood that physical systems enter into naturally entangled states
without the intervention of external controls. As it evolves in time, a chaotic system will visit every part of its phase space, but when its dynamics
have brought it close to a cupolet or to an UPO, the system will track that orbit for an extended period of time. Chaotic entanglement can be regarded
as an entropy-reversing event, and if occurring in a large enough ensemble of chaotic systems, the result could be spontaneous entanglement that
arises naturally and without the influence of an external controller. In theory, this could set off a chain reaction of stabilizations, and it would
be interesting to see if any resulting lattice of entangled cupolets could evoke connections to Ising models~\cite{Karthik2007}.

In the preliminary investigations to date, the chaotic systems are exchanging information, but they are not being driven by physical forces, so one
future research direction would be to investigate mechanisms by which this entanglement would manifest itself in physical systems. Therefore, there is
much potential for cross-fertilization of this work with other research areas and many of the exchange functions in~\cite{Morena2014b} are inspired by
documented studies of interacting physical systems. Consequently, we are investigating certain Hamiltonian systems that are known to be chaotic, as
well as physical systems where an interaction is defined through a short-range force. Just as interesting is the possibility that chaotic entanglement
may be achievable using entirely new materials, whereby the chaotic properties are found at the molecular or atomic level. Indications of such chaotic
behavior have already been documented in~\cite{Bandyopadhyay2004,Bandyopadhyay2005,Habib2006b}, and if such entanglement can be found and manipulated,
then opportunities would exist for new technologies.

It is hoped that the discussion presented here and in~\cite{Morena2014b} will motivate the derivation of additional exchange functions with direct
applicability to other research areas. Doing so could generalize our results and possibly uncover connections between the statistical and
deterministic descriptions of chaotic dynamics, which could in turn be used to explain the natural entanglement that arises between some physical
systems. In summary, chaotic entanglement may well be a still-undiscovered property of certain physical systems and we hope that this research will
lay the groundwork for the discovery of such chaotically-entangled states.



\end{document}